\documentclass[11pt]{article}
\usepackage{amssymb}
\usepackage{amsmath}
\usepackage{graphicx}
\usepackage{color}

\def\beq{\begin{eqnarray}}    
\def\eeq{\end{eqnarray}}      

\newcommand{\CC}{\Lambda}

\newcommand{\rL}{\rho_{\CC}}

\newcommand{\ba}{\begin{eqnarray}}
\newcommand{\ea}{\end{eqnarray}}
\newcommand{\brr}{\begin{array}}
\newcommand{\err}{\end{array}}
\newcommand{\bc}{\begin{center}}
\newcommand{\ec}{\end{center}}

\newcommand{\be}{\begin{equation}}
\newcommand{\ee}{\end{equation}}

\newcommand{\mysection}[1]{\section{#1}
\renewcommand{\theequation}{\thesection.\arabic{equation}}
\setcounter{equation}{0}}

\newcommand{\mysubsection}[1]{\subsection{#1}
\renewcommand{\theequation}{\thesubsection.\arabic{equation}}
\setcounter{equation}{0}}

\begin{document}

\hyphenation{cos-mo-lo-gists un-na-tu-ral-ly ne-cessa-ry dri-ving
par-ti-cu-lar-ly a-na-ly-sis mo-del mo-dels ex-ten-ding e-xam-ples
ho-we-ver}

\begin{center}

{\LARGE \textbf{Thermodynamical aspects of running vacuum models}} \vskip 2mm

 \vskip 8mm

 \vskip 8mm

\textbf{J. A. S. Lima}

\vskip0.25cm
Departamento de Astronomia, Universidade de
S\~ao Paulo, Rua do Mat\~ao 1226, 05508-900, S\~ao Paulo, Brazil

\vspace{0.25cm}
E-mail: jas.lima@iag.usp.br \vskip2mm

\textbf{Spyros Basilakos}

\vskip0.25cm

Academy of Athens, Research Center for Astronomy \& Applied
  Mathematics, \\ Soranou Efessiou 4, 11-527, Athens, Greece

\vspace{0.25cm}
E-mail: svasil@academyofathens.gr \vskip2mm

\textbf{Joan Sol\`a}

\vskip0.25cm
High Energy Physics Group, Dept. d'Estructura i Constituents de
la Mat\`eria, and Institut de Ci\`encies del Cosmos (ICC), \\ Univ.
de Barcelona, Av. Diagonal 647,
 E-08028 Barcelona, Catalonia, Spain

\vspace{0.25cm}
E-mail: sola@ecm.ub.edu \vskip2mm

\end{center}
\vskip 15mm
=

\begin{abstract}
The thermal history of a large class of running vacuum models
in which the effective
cosmological term is described by a truncated power series
of the Hubble rate, whose
dominant term is $\Lambda (H) \propto H^{n+2}$, is
discussed in detail. Specifically, by assuming that the ultrarelativistic particles produced by the vacuum decay 
emerge into space-time in such a way that its energy density $\rho_r \propto T^{4}$,  the temperature evolution law and the increasing entropy function are analytically calculated.  For the whole class of vacuum models explored here we find
that the primeval value of the
comoving radiation entropy density (associated to effectively
massless particles) starts
from zero and evolves extremely fast until reaching a maximum
near the end of the vacuum
decay phase, where it saturates. The late time conservation of the radiation entropy during the adiabatic FRW  phase also guarantees that the whole class of running vacuum models predicts the
same correct value of the present day entropy, $S_{0} \sim 10^{87-88}$ (in natural units), independently
of the initial conditions.  In addition, by assuming Gibbons-Hawking temperature as an initial condition, we find
that the ratio between the late time and primordial vacuum energy densities is in agreement with
naive estimates from quantum field theory, namely, $\rho_{\Lambda 0}/\rho_{\Lambda I} \sim10^{-123}$.
Such results are independent on the power $n$ and suggests that the observed Universe may
evolve smoothly between two extreme, unstable, nonsingular de Sitter phases.

\vskip0.2cm

\end{abstract}

\newpage

\mysection{Introduction}

A non-singular early de Sitter phase driven by a
decaying vacuum energy density was phenomenologically proposed long \phantom{} ago to solve some
problems of the Big-Bang cosmology \cite{LM94, LT96}. The basic idea is closely related to early attempts \phantom{} aimed at solving (or at least alleviating) some cosmic mysteries, such as  the  ``graceful exit" problem, which plague many inflationary scenarios
(for recent reviews see \cite{GKN,Linderev}), and also the CCP or
cosmological constant problem \cite{SW89,SA00,PAD,PR03,Lima2004,CT06,Sola2013}, \phantom{}probably,
the  deepest conundrum of all inflationary theories describing the very early Universe.

Nevertheless, new theoretical developments are suggesting a possible way to circumvent such problems. \phantom{} Results based on the  renormalization group (RG) theoretical techniques of quantum field theory (QFT) in curved spacetimes combined with some phenomenological insights provided a set of
dynamical $\Lambda$(H)-models (or running vacuum cosmologies) described by an even power series of
the Hubble rate \cite{SS02,S08,SS09} (cf. \cite{Sola2013} for a review). \phantom{} In this line, we have discussed in a series of recent works a unified class of models accounting for a complete cosmological history evolving
between two extreme (primeval and late time) de Sitter phases whose
spacetime dynamics is supported  by a dynamical decaying (or running) vacuum energy density
\cite{LBS13,BLS13,P2013,BLS14,LBS15,SolaGomez2015,SolaGRF2015}. {\phantom{} In such models, the effective vacuum energy density is a
truncated power series of the Hubble rate, whose dominant term is $\Lambda (H)  \propto H^{n+2}$, where for the sake of generality the power $n >0$.}
Unlike several inflationary variants endowed with a preadiabatic phase, this decaying vacuum is
responsible for an increasing entropy evolution since the very early Universe, described by the primeval {\phantom{}nonsingular de Sitter spacetime}.

{\phantom{} Several theoretical and observational properties of this large class of nonsingular running vaccum scenarios have been discussed in the literature. In particular, Mimoso and Pav\'on shown their thermodynamic consistency based on the generalized second law of thermodynamics (GSLT) by taking into account quantum corrections to the Bekenstein-Hawking entropy \cite{MP2013}. Many details regarding the late-time dynamics can be found in Refs.\cite{B04,BPS09,B09,SBP2011,BP12}, and especially in the recent, updated and very  comprehensive works \cite{GVS1} and \cite{GVS2}. More recently, even the solution of the coincidence problem has been detailed discussed in the present framework \cite{ZSL2015}, as well as in generic decaying vacuum cosmologies \cite{GSS2006}.}

\phantom{}Finally, let us mention that the running vacuum models under study
have recently been tested against the wealth of
SNIa+BAO+H(z)+LSS+BBN+CMB data -- see\,\cite{MG14} for
a short review -- and they turn out to provide
a quality fit that is significantly better than the $\CC$CDM.
This fact has become most evident in
the  recent works \cite{4sigma,3sigma}, where it is shown that the significance of the fit improvement is at $\sim 4\sigma$ c.l. Therefore, there is plenty of
motivation for further investigating these running vacuum
models from different perspectives, with the hope
of finding possible connections with fundamental aspects
of the cosmic evolution

In the present work,  we focus our attention on  the entropy of the cosmic microwave background radiation (CMBR) generated by this large class of non-singular decaying vacuum cosmology. By considering that the decaying vacuum  process occurs under adiabatic conditions [in the sense that the specific entropy (per particle) is preserved] this means that the radiation produced satisfies the standard scaling laws, namely,  $n_r \propto T^{3}$ and $\rho_r \propto T^{4}$ \cite{Lima96a}.  Under these conditions, the 
final value of the entropy produced by the decaying vacuum supporting the unstable
primeval de Sitter phase is exactly 
the present radiation entropy existing within the current
Hubble radius. Within this framework, the early decaying vacuum process is not plagued with
``graceful exit" problem of most inflationary variants
and generates the correct number,
$S_{0} \simeq 10^{88}$ \cite{KT90},
regardless of the power $n$ present in the
phenomenological decaying $\Lambda$ (H)-term. In addition,  the ratio
between the primeval and the present day vacuum energy densities is
$\rho_{\Lambda 0}/\rho_{\Lambda I}  \simeq 10^{-123}$, as required by some
naive estimates from quantum field theory.

{\phantom{} The article is structured in the following manner.} In section 2 we  justify
 the phenomenological decaying vacuum law adopted in the paper,  whereas in section 3 we
set up the basic set of equations and the transition from the early de Sitter to the
radiation phase is addressed.  {\phantom{} How inflation ends and the temperature evolution law are presented
in section 4,  while in section 5 we discuss the entropy production generated by the decaying vacuum
medium.} Finally, the main conclusions are summarized in section 6.

\mysection{General model for a complete cosmic history}

The general $\Lambda(H)$-scenario accounting for a complete
chronology of the universe (from de Sitter to de Sitter) is based on
the following expression for the dynamical cosmological term defining the relevant class of running vacuum models under consideration \cite{LBS13,BLS13,P2013}:
\begin{equation}\label{powerH}
8\pi G\rho_{\Lambda}\equiv\Lambda(H)= c_0+3\nu H^2 +3\alpha\frac{H^{n+2}}{H_I^n}\,.
\end{equation}
Here $H\equiv\dot a/a$ is the Hubble rate, $a=a(t)$ is the scale factor, and the over-dot denotes derivative with respect to the cosmic time $t$. By definition
$\rL(H)=\CC(H)/8\pi\,G$ is the corresponding vacuum energy density
($G$ being Newton's constant). The even powers of $H$ (therefore $n=2,4,...$) are thought to be of more fundamental origin
due the general covariance of the effective action, as required by
the QFT treatment in curved spacetime \cite{SS02,S08,SS09,Sola2013}. In the numerical analysis, however, we will explore also the cases $n=1$, $3$ and $4$ for comparison.

{\phantom{} The dimensionless free parameters $\alpha$ and $\nu$ have distinct status. The first can be absorbed (for each value of $n$ in the arbitrary scale $H_I$ so that it can be fixed to unity without loss of generality (\phantom{}if the scale of inflation is not precisely known) \cite{SolaGomez2015,ZSL2015}, whereas $\nu$ has been determined from observations
based on a joint likelihood analysis involving SNe Ia, Baryonic Acoustic Oscillations (BAO) and Cosmic Background Radiation (CMB) data, with the result
$|\nu| \equiv {\cal O}(10^{-3})$ \cite{B04,BPS09,B09,SBP2011,BP12,GVS1,GVS2} -- \phantom{} see especially the most recent analyses in which the $\nu=0$ result (associated to the $\Lambda$CDM in the post inflationary time) is excluded at $\sim 4\sigma$ c.l. \cite{4sigma,3sigma}. The small value of $\nu$ is natural since  at late times, the
dynamical model of the vacuum energy cannot depart too much from $\CC$CDM. In this connection, by using a generic Grand Unified Theory (GUT), it has been shown
that $|\nu| \sim 10^{-6}-10^{-3}$ \cite{S08}. Finally, the constant $c_0$ with  the same dimension of $\Lambda$ yields the dominant term
at very low energies, when $H$ approaches the measured value $H_0$ (from
now on the index ``0" denotes the present day values of the quantities).}


\mysection{Basic equations of the $\Lambda(H$) model}

It is well known that the Einstein field equations (EFE) are valid
either for a strictly constant $\CC$ or a dynamical one \cite{LBS13,P2013}. Therefore, using the
vacuum energy density $\rho_{\Lambda}=\Lambda/(8\pi G)$ and the nominal
pressure law $p_{\Lambda}=-\rho_{\Lambda}$ one can write
the EFE in the framework of
a spatially flat FLRW (Friedmann-Lema\^\i tre-Robertson-Walker) space-time
\begin{equation}
R_{\mu \nu}-\frac{1}{2}g_{\mu \nu}R=8\pi G{\tilde T}_{\mu \nu}
\end{equation}
where $R$ is the Ricci scalar and
${\tilde T}_{\mu \nu}=  (p_{m}+\rho_{m})u_{\mu}u_{\nu} - (p_{m}-\rho_{\Lambda})g_{\mu \nu}$
is the total energy-momentum tensor
and the index $m$ refers to the dominant
fluid component (nonrelativistic matter or radiation).
Obviously, in our case, the only difference with
respect to the more conventional field equations is the fact that
$\CC=\CC(H)$. In this framework, the local energy-conservation law
$\nabla^{\mu}\,{\tilde{T}}_{\mu\nu}=0$ which insures the
covariance of the theory {\phantom{} reads:
\begin{equation}
\dot{\rho_{\Lambda}} + \dot{\rho}_{m}+3(1+\omega)\rho_{m}H =0\,, \label{ECL}
\end{equation}
where we have used $p_m = \omega \rho_m$ for the ordinary cosmic fluid}, namely
$\omega=0$ for dust and $\omega=1/3$ for radiation.
In this enlarged framework, the Friedmann equations are given by
\begin{eqnarray}
&&8\pi G\rho_{\rm T}\equiv 8\pi G\rho_m + \Lambda(H)=3H^2 \,,\label{EE}\\
&&8\pi G p_{\rm T} \equiv 8\pi G P_m-\Lambda(H)=-2\dot H-3H^2\label{EE2}\,.
\end{eqnarray}
By combining the above equations with the class of vacuum models (\ref{powerH}) one obtains {\phantom{} the equation driving the evolution of the Hubble parameter}:

\begin{equation}
\label{HE}
\dot{H}= - \frac{3}{2}(1+\omega)H^2\left(1-\nu-\frac{c_0}{3H^2}-\alpha\frac{H^{n}}{H_I^n}\right)\,.
\end{equation}
 A solution of this equation in the high energy regime [where the term $c_0/H^2\ll 1$ of
(\ref{HE}) can be neglected] is given by:
\begin{equation}\label{HS1}
 H(a)=\frac{\tilde H_I}{\left[1+D\,a^{3(1 + \omega)\,n\,(1-\nu)/2}\right]^{1/n}}\,,
\end{equation}
where ${\tilde H_I}=({\frac {1-\nu}{\alpha}})^{1/n} H_I$.

The combination of the EFE
and the expression for $\Lambda$ yields:

\begin{equation}\label{eq:rLa}
  \rho_\Lambda(a)=
\tilde{\rho}_I\,\frac{1+\nu\,D\,a^{3n(1+\omega)(1-\nu)/2}}
{\left[1+D\,a^{3n(1+\omega)(1-\nu)/2}\right]^{1+2/n}}\,,
\end{equation}
\begin{equation}\label{rho_1}
 \rho_r(a)=\tilde{\rho}_I\,\frac{(1-\nu)D\,a^{3n(1+\omega)(1-\nu)/2}}
{\left[1+D\,a^{3n(1+\omega)(1-\nu)/2}\right]^{1+2/n}}\,,
\end{equation}
\begin{equation}
 \rho_{\rm T}(a)=\tilde{\rho}_I\,\frac{1}{\left[1+D\,a^{3n(1+\omega)(1-\nu)/2}\right]^{2/n}}\,.
\end{equation}
The quantity $\tilde{\rho}_I$ in the above equations is the critical energy density
defining the primeval de Sitter stage
\begin{equation}\label{eq:rhoItild}
\tilde{\rho}_I\equiv\frac{3\tilde H_I^2}{8\pi G}.
\end{equation}


\mysubsection{From initial de Sitter stage to radiation phase}\label{sect:initialdeS}
For $\omega = 1/3$ the energy density for
the vacuum and radiation read:
\begin{equation}\label{eq:rLa}
  \rho_\Lambda(a)=\tilde{\rho}_I\,\frac{1+\nu\,D\,a^{2n(1-\nu)}}{\left[1+D\,a^{2n(1-\nu)}\right]^{1+2/n}}\,,
\end{equation}
\begin{equation}\label{rho_1}
 \rho_r(a)=\tilde{\rho}_I\,\frac{(1-\nu)D\,a^{2n(1-\nu)}}{\left[1+D\,a^{2n(1-\nu)}\right]^{1+2/n}}\,,
\end{equation}
We can see from (\ref{eq:rLa})
that the value (\ref{eq:rhoItild}) just provides the vacuum energy density
for $a\to 0$, namely $\rho_{\CC}(0)=\tilde{\rho}_I$. {\phantom{} As $|\nu|\ll 1$ we also see that
$\tilde{\rho}_I /\rho_I\sim\alpha^{-2/n}$, thereby effectively showing that  the constant $\alpha$ can be absorbed in the scale $H_I$, as remarked before.} Let us also emphasize from the previous formula that for $a\to 0$ we have
$\rho_r/\rL\propto a^{2n(1-\nu)}\to 0$, i.e. the very early universe is
indeed vacuum dominated with a negligible amount of radiation. {\phantom{} In the rest of the paper, we neglect the effects proportional to
$\nu$ (which, since it is the coefficient of $H^2$, is not essential for the study of the early
universe, the epoch where the $H^4$-term is fully dominant). Thus, without loss of generality, $H_I$ will be hereafter rescaled so that $\alpha = 1$ and we set $\nu=0$in all the formulae. Within this framework, we have $\tilde H_I = H_I$ and Eq.(\ref{eq:rhoItild}) becomes}

\begin{equation}\label{rhoI}
\tilde{\rho}_I \equiv \rho_I = \frac{3 H_I^2}{8\pi G}\,.
\end{equation}

{\phantom{} In addition,  from Eq.\,(\ref{HS1}) it follows that the scale factor of the universe
takes on an exponential form $a(t)\sim a_i e^{\,H_I\,t}$ as long as the condition
$Da^{3(1+\omega)}\ll 1$ is fulfilled.}  Obviously, this means that
the universe is initially driven by a pure nonsingular de Sitter vacuum state,
and therefore is inflating.
However, the above mentioned  de-Sitter inflationary phase is only
ephemeral. Indeed it is easy to check that in
the post-inflationary regime, i.e. for $a\gg a_i$  (with
$Da^{3n(1+\omega)}\gg 1$), we are led to $H\propto a^{-2}$
(or $a\propto t^{1/2}$) for $\omega = 1/3$ [see Eq. (\ref{HS1})].
{\phantom{} Therefore,  the present model evolves smoothly from
inflation towards the conventional  radiation stage,  thereby  insuring that the initial very large amount of vacuum energy density does not preclude the standard picture of the primordial big-bang nucleossynthesis.

{On the other hand, using equation (\ref{HE})
one may check that the decelerating parameter, $q=-{\ddot
a}/aH^{2}=-1-\dot{H}/H^2$, for arbitrary values of the power index $n$, reads:
\begin{equation}
 q(H) = 2 \left[ 1 - \left(\frac{H}{{H}_I}\right)^{n} \right] - 1 \,, \label{eq3}
\end{equation}
Naturally, the existence of the radiation stage is not enough  to identify  precisely the end of inflation since the deceleration parameter varied from $q=-1$ (de Sitter) to $q=1$ (radiation) and the inflationary period finished when $q=0$, or equivalently, $\ddot a=0$. We will discuss this point in the next section.}

\begin{figure}[t]
  \begin{center}
      \resizebox{0.7\textwidth}{!}{\includegraphics{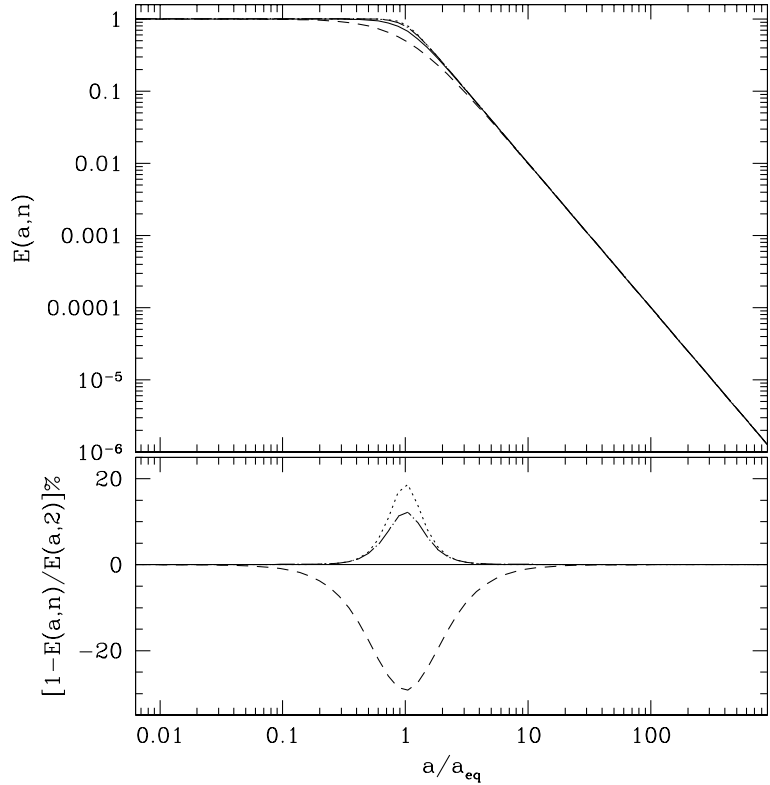}}
      \end{center}
    \caption{{\it Upper Panel:}
Evolution of the normalized Hubble parameter $E(a,n)=H(a,n)/H_{I}$ during the
inflationary epoch and its transition
into the FLRW radiation era. The Hubble parameter is
normalized with respect to $H_I$, and the
scale factor with respect to $a_{eq}$, the value for which
$\rho_{\Lambda} = \rho_r$ (see the text).
The lines correspond to the following $\Lambda(H) \propto H^{n+2}/H^{n}_{I}$
scenarios [see Eq.(\ref{powerH}], namely
$n=1$ (dashed), $n=2$ (solid), $n=3$ ({dot-dashed}) and
$n=4$ (dotted). {\it Lower Panel:} We provide the relative deviation
$[1-E(a,n)/E(a,2)]\%$ of the normalized Hubble parameter for the
$n=1,3,4$ vacuum models with respect to $n=2$.}
  \label{fig1}
\end{figure}


\mysubsection{When exactly Inflation Ends?}

{\phantom{} In order to answer this question we first combine
Eqs. (\ref{eq:rLa}) and (\ref{rho_1}) so as to obtain
the ratio of the radiation energy
density ($\omega=1/3$) to the vacuum energy
density:
\begin{equation}
 \frac{\rho_{r}(a)}{\rho_{\Lambda}(a)} = D a^{2n}\,. \label{eq6}
\end{equation}
In principle, inflation must end when both components\phantom{} - the decreasing vacuum energy density and the created radiation energy density - contribute alike. Assuming that the scale factor at the point of
vacuum-radiation ``equality'' is $a=a_{eq}$  the above expression implies that
$D\,a_{eq}^{2n} = 1$. This relation} enables us to rewrite
the Hubble parameter (\ref{HS1}) in the following way:
\begin{equation}
\label{eq7a}
H(a,n) = \frac{{H}_I}{\left[1 + \left({a}/{a_{eq}}\right)^{2n}\right]^{1/n}}\,.
\end{equation}
It follows that the value of the Hubble function at the
vacuum-radiation equality depends on the value
of the parameter $n$ and the initial scale $H_I$:
\begin{equation}
H_{eq} \equiv H(a_{eq})= \frac{{H}_I}{{2}^{1/n}}. \label{eq7b}
\end{equation}

{ Now, by inserting  this value of $H_{eq}$ into Eq. (\ref{eq3})  we obtain effectively  that $q=0$ as should be expected. Hence, once the arbitrary scale $H_I$ is fixed, the energy scale or the moment for which the inflation ends is readily defined.}

{\phantom{} As remarked before, the start of the radiation phase in
this model is not characterized by the canonical radiation value
$q=1$, as one might have naively expected. \phantom{} Due to the
continuous energy exchange between vacuum and radiation  there is a short period of time in which $q$ goes from
$q = 0$ to the standard result $q = 1$.}
In the begin of the standard adiabatic radiation regime
the deceleration parameter is written as
\begin{equation}
 q(H_{rad}) = 2 \left[1 - \left(\frac{H_{rad}}{{H}_I}\right)^{n} \right] - 1\,.  \label{eq15}
\end{equation}
{\phantom{} In the approach  to the
the radiation phase one may safely assume that it started when $q(H_{rad}) \sim 0.9999$ with
the Hubble parameter given by
$H_{rad}=H_I/(2\times 10^{4})^{1/n}$ [see Eq.(\ref{eq15})], and from Eq.(\ref{eq7a}) we
obtain $a_{rad}/a_{eq} \simeq (2\times 10^{4})^{1/2n}$.


At this point it is appropriate to make the following comments
concerning the cosmic history. First, in the full radiation era
the value of the scale factor of the universe $a_{rad}$
becomes at least one order of magnitude
larger than the corresponding value at the vacuum-radiation
equality. Actually,  the total entropy
generated by the vacuum decaying process does not depend on the  exact value of the ratio
$a_{rad}/a_{eq}$.  Second, when the radiation epoch is well left behind
the Universe goes into the cold dark matter dominated era
[Einstein-de Sitter, $a(t)\propto t^{2/3}$]; and,
after some billion years ($\sim 7$Gyrs)
it enters the present vacuum dominated phase, in which
$\Lambda \simeq {\tilde \Lambda}=$const -- confer \cite{BPS09,GVS1} and \cite{4sigma,3sigma}.

In the upper panel of Fig.1 we provide the evolution of the
normalized Hubble parameter
$E(a,n)=H(a,n)/H_{I}$ for the following vacuum models
$n=1$ (dashed line), $n=2$ (solid line), $n=3$ (long dashed line) and
$n=4$ (dotted line) [see Eq.(\ref{powerH})]. We observe
that for $a \ll a_{eq}$ the cosmic evolution begins
from an unstable inflationary phase [early de Sitter era,
$H \simeq H_I$] powered by the huge value $H_I$
presumably connected to the scale of a Grand Unified Theory (GUT).
Obviously, when the primeval inflationary era is left behind, particularly
for $a \gg a_{eq}$, the cosmic evolution enters smoothly in
the standard radiation period $H\propto a^{-2}$.
Overall, we would like to emphasize
that the above natural mechanism for graceful exit is universal for the whole
class of vacuum models which obey the restriction $n > 0$.
Subsequently when the $c_{0}/H^{2}$ quantity in Eq.(\ref{HE}) starts
to dominate over $H^{n}/H^{n}_{I}$ (where $H\ll H_{I}$)
the radiation component becomes sub-dominant and
the matter dominated era appears. This implies that Eq.(\ref{powerH})
reduces to $\Lambda(H)={\tilde \Lambda}+3\nu (H^{2}-H^{2}_{0})$ which
generalizes the traditional $\Lambda$CDM model. In this case
the vacuum at the present time (cosmological constant)
becomes ${\tilde \Lambda}=3c_{0}+3\nu H^{2}_{0}$, where $H_{0}$ is the Hubble
constant. More details regarding the late-time dynamics can be amply found in Refs.\cite{B04,BPS09,SBP2011,BP12}, and especially in the recent, updated and very  comprehensive works \cite{GVS1} and \cite{GVS2}.

Finally, by using $n=2$ (corresponding to $\Lambda(H) \propto H^{4}/H^{2}_{I}$ in the early universe)
as a fiducial model in the large class (\ref{powerH}), we can appreciate in the bottom panel of Fig.1 the relative deviation of the normalized Hubble parameters
$E(a,n)$ with respect to the $n=2$ solution $E(a,2)$.
Obviously, when we are far away from the epoch of the
vacuum-radiation equality
the deviations from the fiducial $E(a,2)$ case are extremely small. On the other
hand there is a visible deviation from the latter around the transition region
$a\to a_{eq}$. This deviation becomes at the level of $\sim -30\%$
and $\sim +20\%$ for $n=1$ and $n>2$ respectively.


\mysection{Radiation and its Temperature Evolution Law}

Assuming an ``adiabatic'' decaying vacuum  it has been
found that the specific entropy of the  produced massless
particles remains constant, despite the fact that the total entropy can be
increasing\footnote{More precisely, the constancy of
the specific entropy (per particle) of the produced particles defines
the ``adiabatic" vacuum decaying process (see Refs.\cite{Lima96a,Lima97}).}. {\phantom{} This implies that}  the energy and
the number density as {\phantom{} a function of the temperature are given by the
the standard expressions}, namely:
$\rho_r \propto T_r^{4}$ and $n_r \propto T_r^{3}$, but
the temperature does not obey the scaling relation
$T_r(t) \sim a(t)^{-1}$. {\phantom{} Such results were firstly derived based on a covariant nonequilibrium thermodynamic description \cite{Lima96a,Lima97}), and, more recently, using a kinetic theoretic approach \cite{LB14}. In what follows, we discuss the temperature evolution law in the present framework along these lines.}

Let us start with Eq.(\ref{rho_1}) which is rewritten in terms
of $a_{eq}$ as follows
\begin{equation}\label{eq:rhorrhoII}
\rho_{r} ={\rho}_I \frac{\left(a/a_{eq}\right)^{2n}}{\left[1 +
(a/a_{eq})^{2n}\right]^{1+2/n}}=\frac{\pi^2}{30} g_{\ast} T_{r}^{4}\,,
\end{equation}
{\phantom{} where in the last equality we included  the degrees of freedom (d.o.f.) of the created massless modes
through the  $g_{*}$-factor (see \cite{KT90}). Now, by solving for the temperature we find:}
\begin{equation}\label{eq:Tr}
T_r = {2}^{\frac{n+2}{4n}}T_{eq}\frac{(a/a_{eq})^{n/2}}{\left[1 +
(a/a_{eq})^{2n}\right]^{\frac{n+2}{4n}}}\,,
\end{equation}
where $T_{eq}=T_r (a_{eq})$. Obviously, $T_{eq}$ is the
maximum value of the radiation temperature (\ref{eq:Tr}) which is defined by
\begin{equation}\label{Temp}
T_{eq} = \left(\frac{15\,\rho_I}{2\pi^{2}g_*}\right)^{1/4} =
\left(\frac{45\,H_I^{2}}{16\pi^{3}Gg_*}\right)^{1/4}\,.
\end{equation}
As it is expected $T_{eq}$ is given in terms of the
arbitrary initial scale of the primeval
de Sitter phase ($\rho_I$, or equivalently $H_I$). {\phantom{} Unlike the value of $H_{eq}$ (see Eq.(\ref{eq7b})), this value of $T_{eq}$ is valid for the whole class of models since is does not depend on the power $n$.  In particular, this unique maximum temperature suggests that the total entropy generated within the horizon and presently observed is basically the same for all models (see next section).}

In the top panel of  Fig.\,2 we present the temperature evolution
for several values of $n$. { Note that in the very beginning of the evolution (when $a \mapsto 0$), the temperature of the created photons is also zero in accordance with the fact that $\rho_r = 0$ (see Eq. (\ref{eq:rhorrhoII})). However, for finite values of $a \neq 0$,  we observe the existence of two regimes.} In the first, $a \ll a_{eq}$, the radiation
temperature (\ref{eq:Tr}) increases as $T_r\propto a^{n/2}$
(it is linear for $n=2$ \cite{LBS15}), reaching of course
its maximum value at $a=a_{eq}$. {\phantom{} In this non-adiabatic regime, the vacuum instability \phantom{} guides
constantly the model to the standard radiation epoch \phantom{} from a process that started in the non-singular de Sitter phase. Note that the evolution is different from inflationary models where  a highly non-adiabatic ``reheating''
process happens immediately after the adiabatic evolution of the inflaton field
(see \cite{KT90} and references therein).
In the opposite regime,  $a \gg a_{eq}$, we are well within the radiation epoch when all running vacuum models
decrease in the classical way, that is,  following an adiabatic scaling law, $T_r
\propto a^{-1}$.}  As we shall discuss bellow, the power-law increasing of the
temperature up to the vacuum-radiation equality, beyond of which
the universe enters in the standard temperature regime, is
the reason of the large radiation entropy observed in the present epoch.

For the purpose of the present study it is also
important to find a way to calculate the
primeval parameters ($\rho_I, H_I$) from first principles
and indeed one may use the following approaches
(from now on we consider natural units, $\hbar=k_B=c=1$).
A first possibility is to consider that the
initial de Sitter energy density $\rho_I$ is related
with the Planck scale  $M_P$ through $\rho_I= M_{P}^{4}$.
An alternative approach is to connect $\rho_I$ to the GUT energy scale
$M_X\sim 10^{16}$ GeV and thus $\rho_I=M_X^{4}$
(see Refs. \cite{SolaGomez2015,SolaGRF2015}). A third possibility is to use
the event horizon (EH) of the de Sitter space-time.
From the analysis of Gibbons \& Hawking \cite{GH77} we know that
the temperature of the de Sitter EH in natural units
is ${T}_{GH}={{H}_I}/{2\pi}$, where
$H_I$ is the (constant) Hubble parameter at the de Sitter epoch.
In our case, following the third approach and {\phantom{} combining the last equality of Eq.\,(\ref{eq:rhorrhoII}) and Eq. (\ref{rhoI})} we obtain
%
\begin{equation}\label{GibbHI}
T_I = \frac{{H}_I}{2\pi} =\left(\frac{45}{\pi g_*}\right)^{1/2}\,M_{P}\,,\,\,\,\, \ \ \
\ \,\,\,\,\rho_I = \frac{135}{2g*}M_{P}^{4}\,.
\end{equation}
Then inserting the above expressions into Eq.(\ref{Temp}) it is easy
to show that the characteristic radiation temperature $T_{eq}$
becomes
\begin{equation}\label{GibbTI}
 {T}_{eq}=\left(\frac{45}{2\pi\,g_*}\right)^{1/2}M_{P}\,.
\end{equation}
It should be stressed that all the above
characteristic scales are slightly below the
corresponding Planck scale, which means that
we are inside in the semi-classical QFT regime.
For instance, by taking $g_{*}=106.75$ which
corresponds to the particle content of the
standard model of particle physics we find from Eq.(\ref{GibbTI}) that
${T}_{eq}\simeq 0.26\, M_P$  which as expected does not depend on the power $n$. In this respect if we take into account the
number of light d.o.f in the GUT then we find that the characteristic temperature  is still smaller (nearly  $10\%$ of Planck mass). { At this point it should be stressed  that the primordial Gibbons-Hawking thermal bath was used only as a peculiar initial condition to fix the arbitrary scale $H_I$. As we shall see in the next section, the ratio between the  very early and late time vacuum energy densities depends only on the pair ($H_I, H_F$) characterizing the extreme de Sitter phases. It has the expected magnitude thereby also contributing to alleviate the so-called cosmological constant problem in the context of the such models (see Eq. (\ref{eq:ALLE})).}

\mysection{The Radiation Entropy}
The total entropy of the radiation included in the present
Hubble radius ($d_H\simeq H_0^{-1}$) reads:

\begin{equation}\label{eq:TotalEntropyToday}
S_{0} = \frac{2\pi^2}{45}\,g_{s,0}\,T_{r0}^3\,H_0^{-3}\simeq 2.3
h^{-3} 10^{87}\sim 10^{87-88}\,,
\end{equation}
where $T_{r0}\simeq 2.725$$^{\circ}$K $\simeq 2.35\times 10^{-13}$
GeV is the CMB temperature at the present time  and $H_0=2.133\,h\times 10^{-42}$
GeV (with $h\simeq 0.67$) is the present day Hubble parameter\footnote{Following standard lines, we have assumed the
coefficient $g_{s,0}=2+6\times
(7/8)\left(T_{\nu,0}/T_{r0}\right)^3\simeq 3.91$ is the entropy
factor for the light d.o.f. at the present epoch in
which we have used $T_{\nu,0}/T_{r0}=\left(4/11\right)^{1/3}$ \cite{KT90}}.

\begin{figure}[t]
  \begin{center}
      \resizebox{0.7\textwidth}{!}{\includegraphics{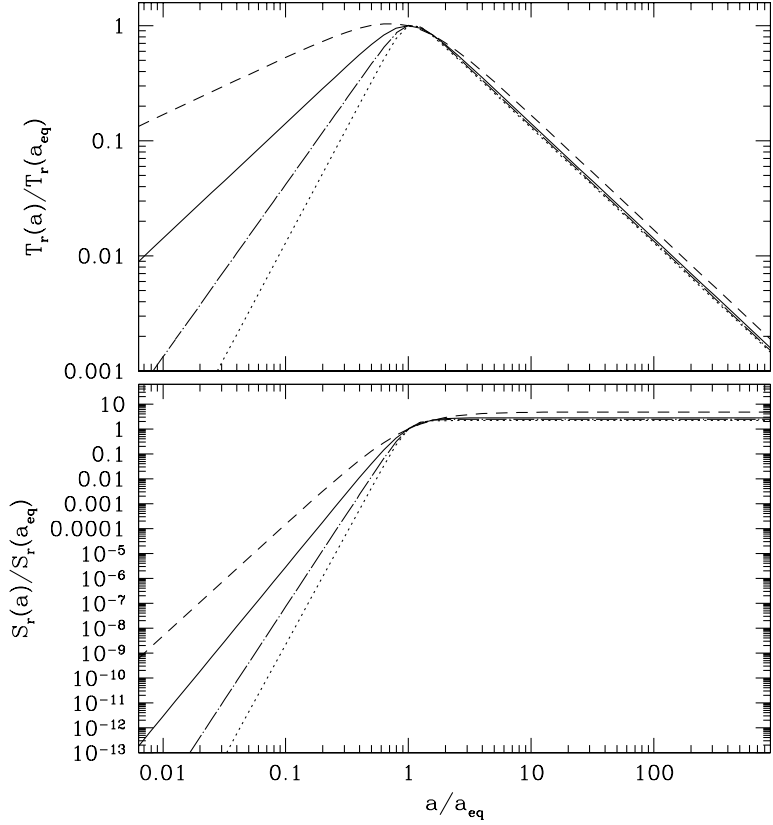}}
      \end{center}
    \caption{{\em Top Panel}: The
evolution of the radiation temperature (normalized with respect to
its maximum value) during the inflationary period
for several values of the free parameter $n$
(see caption of Fig.1 for definitions) and its transition
into the FLRW radiation era. {As in Fig.1, the lines correspond to the following scenarios
$n=1$ (dashed), $n=2$ (solid), $n=3$ (dot-dashed) and
$n=4$ (dotted)}.
{\em Bottom Panel}: The evolution of the
normalized comoving entropy from the inflationary period (where it
increases) until reaching the saturation plateau for $a/a_{rad} \geq
1$. The asymptotic value corresponds the total entropy at present.}
  \label{fig2}
\end{figure}

In order to check the viability of our model we need to take into account
that the total entropy should be measured from the initial entropy
generated by the decaying vacuum. Since in our vacuum model
the BBN proceeds fully standard,
the equilibrium entropy formula remains valid because
only the temperature law is modified \cite{Lima96a,Lima97,LB14}. Therefore,
the radiation entropy per comoving volume is given by the well known expressions:
\begin{equation}
S_{r} \equiv \left(\frac{\rho_{r} + p_{r}}{T_{r}}  \right) a^3 \equiv
\frac{4}{3} \frac{\rho_{r}}{T_r} a^3= \frac{2\pi^2}{45} g_{s}
T_{r}^{3} a^3\,,\label{eq19}
\end{equation}
{\phantom{} where $g_s$ is the entropy factor at temperature
$T_r$ (at very high temperature $g_s$ is
essentially equal to the effective number of massless species,
$g_{*}$.} However, for lower values there is a correction related to the
freeze out of neutrinos and electron-positron annihilation).


With the help of equation  (\ref{eq:Tr}) the comoving entropy (\ref{eq19}) can be expressed as a function of the scale factor as follows:
\begin{equation}
S_{r} = \frac{2^{(7n+6)/4n}\pi^2g_{s}}{45}\,T_{eq}^{3}a_{eq}^{3} f_n(r)\,,
\label{eq:Sr}
\end{equation}
where $r\equiv a/a_{eq}$ and the function $f_n(r)$ depends on the parameter $n$ as given below:
\begin{equation}\label{eq:deffr}
f_n(r)=\frac{r^{\frac{3(n+2)}{2}}}{\left(1+r^{2n}\right)^{\frac{3(n+2)}{4n}}}\,.
\end{equation}
The obtained result for the comoving entropy boils down to the one derived in Ref.\cite{LBS15} for $n=2$, as it should.
Note also that $\lim_{a \rightarrow 0} S_{r} = 0$, as it ought to be
expected from the fact that the initial de Sitter state is supported by a pure vacuum (no radiation fluid).
We also see that during the inflationary phase ($a\ll a_{eq}$),  that is, at the early stages of the
evolution, the total comoving radiation entropy of the Universe increases very fast; in fact,
proportional to $a^{3(1+n/2)}$. For instance, for $n=2$ (corresponding to $H^4$-driven inflation)
the initial entropy raises as $\sim a^6$. Note also that for $a=a_{eq}$,
$f_n(r=1)=1/8^{(n+2)/4n}$ and the associated value $S(a_{eq})$ is still not
the total comoving  entropy that the decaying vacuum is able to
generate (see discussion below Eq. (\ref{eq3})). This occurs only
when $a=a_{rad}$ so that $r^{2n}=(a_{rad}/a_{eq})^{2n} \gg 1$ and
$f_n(r) \simeq 1$ for all practical purposes. {\phantom{} At this point  the
generated comoving entropy $S_r$ reaches the final value,
$S^{f}_{rad}=S_r(a_{rad})$, when the standard radiation phase begins.}

It thus follows that the asymptotic ({\it adiabatic}) value  of
(\ref{eq:Sr}), defined by $f(r) \simeq 1$ for $r\gg1$, is given by:
\begin{equation}\label{eq:Srad1}
S_{r}\to S_{rad}^f=
\left(\frac{2^{(7n+6)/4n}\pi^2g_{s}}{45}\right)T_{eq}^{3}a_{eq}^{3}\,,
\end{equation}
{\phantom{} a saturated value that must be compared with the present day entropy since the subsequent evolution of the model is isentropic}


In the bottom panel of Fig.\,2 we show
the entropy as a function of
the ratio $r=a/a_{eq}$ for several values of $n$.
Notice, that the entropy is scaled to its value at the
vacuum-radiation equality.
Initially, for $a\ll a_{eq}$ the amount of entropy
differs among the vacuum models but when $a\to a_{eq}$ the corresponding
entropies start to converge and subsequently they reach
a plateau, namely $S_r(a)/S_r(a_{eq})\to 8^{(n+2)/4n}$ for $a\gg a_{eq}$
that characterizes the standard
adiabatic phase, which is sustained until the present days because
the bulk of the vacuum energy $\Lambda(H)\propto H^{n+2}/H^{n}_{I}$
already decayed.

Armed with the above expressions we now compute the prediction of the total
entropy inside the current horizon volume $\sim H_0^{-3}$. Using the temperature
evolution law [see Eq. (\ref{eq:Tr})] one may see that the expression
$T_{eq}^{3}a_{eq}^{3}$ which appears in the final entropy as given by
(\ref{eq:Srad1}) is equal to $2^{-3(n+2)/4n}T_{rad}^{3}a_{rad}^{3}$, where
$T_{rad}=T_r(a_{rad})$ and $a_{rad}$
was defined in Sect.\,\ref{sect:initialdeS}
[see discussion below Eq. (\ref{eq15})]. Note that the $n$-dependence cancels out and we arrive at to the same final result (the power $n$ is important only in the inflationary phase since it determines  the time scale in which the radiation equilibrium phase is attained):


\begin{equation}\label{eq:Srad}
S_{rad}^f=\left(\frac{2\pi^2g_{s}}{45}\right){T_{rad}^{3}\,a_{rad}^{3}}=
S_{0}\,,
\end{equation}
where $S_{0}$ is given by (\ref{eq:TotalEntropyToday}). In the last
step we used the entropy conservation law of the standard adiabatic
radiation phase, which implies that $g_s\,T_{rad}^{3}\,a_{rad}^{3}=
g_{s,0}\,T_{r0}^3\,a_0^{3}$.

{\phantom{} It should be stressed that in the \phantom{}very early de Sitter era, the radiation
entropy is zero. However, it increases steadily as \phantom{}$S_r\sim a^{3(1+n/2)}$
and, finally, as shown in the bottom panel Fig. 2, deep inside the radiation \phantom{}stage becomes constant and
approaching its asymptotic present day observed value, $S_0$.}

In other words the running vacuum model
provides an overall past evolution to
the present $\Lambda$CDM cosmology by connecting smoothly between two
extreme cosmic eras (early inflation and dark energy)
driven by the vacuum medium [see Eq.(\ref{HE})]. Specifically,
using Eq.\,(\ref{GibbHI}) we find that
the ratio between the associated vacuum energy densities becomes
\begin{equation}
\label{eq:ALLE}
\frac{\rho_{\Lambda 0}}{\rho_{\Lambda I}} \equiv \frac{\rho_{\Lambda _F}}{\rho_{\Lambda I}}= \frac{H_F^{2}}{H_I^{2}} =
\frac{H_0^{2}\Omega_{\Lambda_0}}{H_I^{2}}=
\frac{g_*}{180\pi}\,\frac{H_0^{2}\Omega_{\Lambda_0}}{M_{P}^{2}} \mapsto \rho_{\Lambda 0} \simeq 10^{-123} \rho_{\Lambda I}\, ,
\end{equation}
in agreement with traditional estimates based on quantum field theory
(see Refs. \cite{SW89} and \cite{PAD}). { Note that the late time de Sitter scale, $H_F = H_0^{2}\Omega_{\Lambda_0}$, was used in the above expression\cite{ZSL2015}.}
To conclude, the decaying vacuum model explains the amount of the
radiation entropy and simultaneously it also alleviates the so-called cosmological constant problem.

Needless to say, for a final resolution of the problem one needs to understand the ultimate origin of the current value of the cosmological constant. Remarkably, the obtained description of the cosmic history is based on a unified dynamical $\Lambda(H)$-term accounting for both the vacuum energy of the early and of the current universe. An alternative approach to such description, based on the Grand Unified Theory framework, can be shown to provide similar results, see \cite{SolaGomez2015,SolaGRF2015}. Somehow this shows that the obtained results are truly robust and independent of the initial conditions reigning in the primeval Universe.

\mysection{Discussion and Conclusions}

In this paper we have addressed a fundamental issue concerning the thermodynamics of the early Universe. It is well-known that within the context of the concordance cosmological model, or $\Lambda$CDM model, the thermal history is incomplete and it leads to inconsistencies with the present observations. One of the main problems is the well-known horizon problem, which can be rephrased thermodynamically as the entropy problem. The concordance model, in fact, cannot offer an explanation for the large entropy enclosed in our Hubble sphere, which is tantamount to say that it cannot explain the large amount of causally disconnected regions contained in it.

The well-known solution to these problems is inflation, a patch that has to be added to the incomplete  $\CC$CDM description. While in the more traditional approach inflation is accomplished by postulating the existence of a new fundamental scalar field called inflaton, in the present work we have proposed an entirely different (but no less efficient) framework. It is based on the properties of a large class of non-singular decaying vacuum models whose structure is that of a truncated power series of the Hubble rate, $\Lambda(H)$. The involved powers to describe inflation must be higher than $H^2$ since the latter can be relevant only for the current Universe.

In this work we have explored an inflationary dynamics {\phantom{} where the decaying vacuum} is triggered by an arbitrary higher power of the Hubble rate,
$H^{n+2}$ ($n > 0$)  recently proposed \cite{LBS13,BLS13,P2013}.

In such unified model of the  vacuum energy, the effective cosmological term is  a dynamical quantity, which evolves extremely fast in the early Universe and goes through an approximate de Sitter phase in our time -- the dark energy epoch. In principle,  the late time $\Lambda(H)$-Universe as described by Eq. (2.1) remains very close to the concordance model, but it stills carries a mild vacuum evolution (hence a mildly evolving cosmological term) compatible with observations. In principle, such a term  may act as a smoking gun of its lengthy and energetic history; {\phantom{} indeed, a much richer history than that associated to the idle $\CC$-term inherent to the $\CC$CDM model. However, since the late time entropy production is very small, for  the sake of simplicity, we have taken the $\nu$ parameter equal to zero. Therefore, the model discussed here can be seen as a primeval nonsingular phase of the standard $\Lambda$CDM model.\phantom{} Once we leave well back the early times, the $\nu$-parameter recovers an important role which, in point of fact, makes the running vacuum models not only compatible with the current cosmological data but fully competitive with the $\Lambda$CDM description\,\cite{4sigma,3sigma}.}


We have studied in detail some \phantom{}important thermodynamical \phantom{}aspects of that class of dynamical vacuum models.
 Most noticeably we have focused on the issue
of the radiation entropy, its origin
and generation in the first stages of the
primeval Universe, and then its final impact on the current
epoch. Our calculations were based on the assumption that the produced radiation from vacuum decay satisfy the standard relations, $n_r \propto T^{3}$ and $\rho_r \propto T^{4}$ \cite{Lima96a}, a hypothesis related with the idea of an ``adiabatic" decaying vacuum and the fact that the specific entropy is preserved during the process \cite{Lima96a}. The basic result is that at early times the temperature of the radiation increases ($T_{r} \propto a^{n/2}$) until its maximum value determined by the equilibrium temperature $T_{eq}$  of the vacuum-radiation transition (see Fig. 2) and the same happens with radiation energy density. As a consequence, the entropy also increases at very early stages ($S_r \propto a^{3(n+2)/2}$)
being later on conserved during the radiation epoch (neglecting the photon entropy produced in the electron-positron annihilation).
In this context we have found that the large
amount of radiation entropy now ($S_0\sim 10^{{87}-{88}}$ in natural units)
can be fully accountable in our dynamical vacuum context.
We have first shown that the entropy was produced during
the inflationary process itself  at
the expense of the continuous vacuum decay. Subsequently, its
production stagnated and this occurred shortly after the vacuum had lost its energetic power and the Universe entered the standard radiation phase. From this point onwards the adiabatic evolution of the cosmos carried the
comoving entropy unscathed until our days.
Overall, the wide class of running
$\Lambda(H)$-vacuum models provides not only an alternative scenario
for inflation (beyond the traditional inflationary scalar field
models, some of them in serious
trouble after the analysis of {\it Planck 2015} data
\cite{Planck2015}),
but also  a new clue for graceful exit, which is indeed fully guaranteed within the $\Lambda(H)$-cosmology context and does not depend on the power $n$ of the Hubble rate. The remarkable independence of both the graceful
exit and the entropy prediction
from the power $n$ singles out such class
of dynamical vacuum models from the rest.

Finally, as originally pointed out, the thermodynamical solution insures that no horizon problem exists
because all points of the current Hubble sphere remained causally
connected as of the early times when the huge entropy was generated
by the decaying dynamics of the primeval vacuum.
Interestingly enough, the above features are not only
universal for the entire class of $\Lambda(H)$-models but also independent of the initial conditions of the early Universe.
They provide a rather robust
basis for the dynamical $\Lambda(H)$-cosmology, which is currently being tested and will continue being tested against the increasingly accurate observations.

\vspace{1cm}


\noindent{\bf Acknowledgments:} \vspace{0.2cm}

JASL is partially supported by the Brazilian Research Agencies, CAPES, CNPq and FAPESP (PROCAD2013, INCT-A and LLAMA, projects).
JS has been supported in part by FPA2013-46570 (MICINN),
CSD2007-00042 (CPAN), 2014-SGR-104 (Generalitat de Catalunya) and MDM-2014-0369 (ICCUB).
SB also acknowledges support by the Research Center for Astronomy of
the Academy of Athens in the context of the program  ``{\it Tracing
the Cosmic Acceleration}''. JASL and SB are also grateful to the
Department ECM (Universitat de Barcelona) for the hospitality and
support when this work was being finished.

\appendix

\section{FRW radiation phase: Wien's law and the relation $T_{rad}/T_I$}

\hskip 0.6cm In this appendix a simple argument based on the equilibrium Wien's law is adopted to show that the decaying vacuum drives the model progressively to the radiation phase without reheating period (no exit problem).

To begin with we recall that the initial conditions in our picture are $T_r = 0$ and $T_{GH} = T_I = H_I/2 \pi$. However, due to the evolution of the Universe, the Hubble parameter and the horizon temperature diminish while  the temperature of the created radiation increases due to the continuous vacuum decay.


The model evolves out of equilibrium because the entropy generation is concomitant with the inflationary process. In principle,  the standard radiation FRW phase is reached when the Wien law becomes strictly valid. In what follows we show that it provides an useful constraint on the value of the temperature in the begin of the FRW phase thereby suggesting the physical consistency of the model.

In natural units, the standard Wien's law  ($\lambda_m T = 0.290\,cm.\,K$) reads:

\begin{equation}
\lambda_m T = 1.27\,\,.
\end{equation}

In order to equalize the horizon temperature, the wavelength at the maximum black body intensity (in the begin of the FRW phase, $H=H_{rad}, T=T_{rad}$) is expected to be $\lambda_m < H_{rad}^{-1}$.  Hence, the Wien's law takes the form:

\begin{equation}
1.27 = \lambda_m T < T_{rad}/H_{rad}.
\end{equation}
Now, in our model we know that $H_{rad} \sim H_I/(2.10^{4})^{1/n}$ (see section 3.2) and $H_I = 2\pi T_I$. Thus, it follows that

\begin{equation}
1.27 < T_{rad}/H_{rad} < \frac {(2.10^{4})^{1/n} T_{rad}}{2\pi T_I},
\end{equation}
which can be rewritten as:

\begin{equation}
\frac{T_{rad}}{T_I} > \frac{7.98}{(2.10^{4})^{1/n}}
\end{equation}
The above relation shows two interesting aspects of the model: (i) $T_{rad}$ is much smaller than $T_{eq}=T_I/2^{1/n}$, but it can assume values not dramatically too low in comparison with the both characteristics scales ($T_I$ and $T_{eq}$); a result in agreement with the demonstrated progressive approach to the FRW phase (see Eq. (\ref{eq:Tr}) and the associated comments), and (ii) it also suggests that there is no exit problem in our model (the reheating process is not needed for this class of decaying vacuum models).  For instance, by taking n= 1, 2 we see that $T_{rad} > 4.10 ^{-4} T_I$, \, and \, $T_{rad} > 0.056T_I$, respectively.  Note also that the  inequality is still more safely satisfied  for values of $n \leq 2$. This is comprehensible because inflation ends faster for higher values of $n$.


\newcommand{\JHEP}[3]{ {JHEP} {#1} (#2)  {#3}}
\newcommand{\NPB}[3]{{ Nucl. Phys. } {\bf B#1} (#2)  {#3}}
\newcommand{\NPPS}[3]{{ Nucl. Phys. Proc. Supp. } {\bf #1} (#2)  {#3}}
\newcommand{\PRD}[3]{{ Phys. Rev. } {\bf D#1} (#2)   {#3}}
\newcommand{\PLB}[3]{{ Phys. Lett. } {\bf B#1} (#2)  {#3}}
\newcommand{\EPJ}[3]{{ Eur. Phys. J } {\bf C#1} (#2)  {#3}}
\newcommand{\PR}[3]{{ Phys. Rep. } {\bf #1} (#2)  {#3}}
\newcommand{\RMP}[3]{{ Rev. Mod. Phys. } {\bf #1} (#2)  {#3}}
\newcommand{\IJMP}[3]{{ Int. J. of Mod. Phys. } {\bf #1} (#2)  {#3}}
\newcommand{\PRL}[3]{{ Phys. Rev. Lett. } {\bf #1} (#2) {#3}}
\newcommand{\ZFP}[3]{{ Zeitsch. f. Physik } {\bf C#1} (#2)  {#3}}
\newcommand{\MPLA}[3]{{ Mod. Phys. Lett. } {\bf A#1} (#2) {#3}}
\newcommand{\CQG}[3]{{ Class. Quant. Grav. } {\bf #1} (#2) {#3}}
\newcommand{\JCAP}[3]{{ JCAP} {\bf#1} (#2)  {#3}}
\newcommand{\APJ}[3]{{ Astrophys. J. } {\bf #1} (#2)  {#3}}
\newcommand{\AMJ}[3]{{ Astronom. J. } {\bf #1} (#2)  {#3}}
\newcommand{\APP}[3]{{ Astropart. Phys. } {\bf #1} (#2)  {#3}}
\newcommand{\AAP}[3]{{ Astron. Astrophys. } {\bf #1} (#2)  {#3}}
\newcommand{\MNRAS}[3]{{ Mon. Not. Roy. Astron. Soc.} {\bf #1} (#2)  {#3}}
\newcommand{\JPA}[3]{{ J. Phys. A: Math. Theor.} {\bf #1} (#2)  {#3}}
\newcommand{\ProgS}[3]{{ Prog. Theor. Phys. Supp.} {\bf #1} (#2)  {#3}}
\newcommand{\APJS}[3]{{ Astrophys. J. Supl.} {\bf #1} (#2)  {#3}}

\newcommand{\Prog}[3]{{ Prog. Theor. Phys.} {\bf #1}  (#2) {#3}}
\newcommand{\IJMPA}[3]{{ Int. J. of Mod. Phys. A} {\bf #1}  {(#2)} {#3}}
\newcommand{\IJMPD}[3]{{ Int. J. of Mod. Phys. D} {\bf #1}  {(#2)} {#3}}
\newcommand{\GRG}[3]{{ Gen. Rel. Grav.} {\bf #1}  {(#2)} {#3}}



\end{document}